\documentstyle[preprint,aps]{revtex}
\begin{document}
\draft
\preprint{ }
\begin{center}
{\bf CORRELATION EFFECTS IN  NUCLEAR TRANSPARENCY}
            \\
\vspace{0.7cm}
L.~L.~Frankfurt
            \\
{\em School of Physics and Astronomy
            \\
Raymond and Beverly Sackler Faculty of Exact Sciences
            \\
 Tel Aviv University, Ramat Aviv 69978, Israel, Tel Aviv, Israel,
            \\
on leave of absence from the St.Petersburg Nuclear Physics
Institute, Russia}
            \\
\vspace{0.5cm}
E.~J.~Moniz
            \\
{\em  Center for Theoretical Physics
            \\
Department of Physics and Laboratory for Nuclear Science
            \\
Massachusetts Institute of Technology
            \\
Cambridge, MA 02139 USA
            \\
and
            \\
Institute for Theoretical Physics III
            \\
 University of Erlangen-N\"urnberg,
            \\
Erlangen, Germany }
            \\
\vspace{0.5cm}
M.~M.~Sargsyan
            \\
{\em School of Physics and Astronomy
            \\
Raymond and Beverly Sackler Faculty of Exact Sciences
            \\
Tel Aviv University, Ramat Aviv 69978, Israel, Tel Aviv,
Israel,     \\
also at the Yerevan Physics Institute, Yerevan, 375036,
Armenia}    \\
\vspace{0.5cm}
 M.~I.~Strikman
            \\
{\em Pennsylvania State University, University Park, PA
16802, USA, \\
also at the St.Petersburg Nuclear Physics Institute,
Russia}     \\
\vspace{0.5cm}

\vspace{3.4cm}
\end{center}
\pagebreak
\vspace{0.7cm}
\begin{abstract}
The Glauber approximation is used to calculate the contribution of
nucleon correlations in high-energy $A(e,e'N)$ reactions. When the
excitation energy of the residual nucleus is small, the increase
of the nuclear transparency due to correlations between the struck
nucleon and the other  nucleons  is mostly compensated by a decrease
of the transparency due to the correlations between non  detected
nucleons. We derive  Glauber model predictions for nuclear transparency
for the differential cross section when  nuclear shell level excitations
are  measured. The role of correlations in color transparency
is briefly discussed.
\end{abstract}


\narrowtext

\section{Introduction}
The semiclassical approximation  improves with increasing collision energy.
This theoretical expectation is supported by the observation that the
Glauber approximation describes quite well the recent experimental data on
high-energy (e,e'p) reactions\cite {NE18}.  Thus this theoretical framework
makes numerous issues in nuclear physics amenable to quantitative study at
new facilities, such as CEBAF or HERMES, and provides a baseline for studying
color coherent phenomena in the collisions of high energy particles with
nuclei.
Such phenomena may provide promising new methods of investigation of the
nonperturbative and perturbative  QCD.  One example is the color transparency
(CT) phenomenon - suppression of the final state interaction (FSI) in  high
energy quasielastic large angle reactions with nuclei. Initial motivation
for the dominance of small size configurations came from the analysis of
the leading perturbative QCD diagrams \cite{MU82,BR82} which should
dominate  at very large momentum transfer. However  the recent  analysis
\cite{FMS93} of   realistic models of a nucleon (pion) has found that the
electromagnetic form factors of a nucleon  and a pion are dominated by
smaller than average size configurations already in the nonperturbative
domain at $Q^2 \ge 2-3 (GeV/c)^2$.  Hence it seems important to study CT
in the $A(e,e'N), A(e,e'NN)$ processes at as small $Q^2$ as possible.
However to achieve these aims one needs both dedicated high resolution
experiments at intermediate $Q^2$ (see for example \cite{NE18,EVA}) and
calculations of nuclear transparency  within the standard Glauber theory.
Such calculations should include nuclear effects such as nucleon
correlations in nuclei and the nuclear shell effects.

We approach these problems by adapting the technique  developed for high
energy hadron scattering off nuclei in the early seventies by Moniz, Nixon
and Walecka \cite{MNW,MN} and by Yennie \cite{YN}. With the Glauber
approximation \cite{GL} it is easily shown that nucleon correlations
in high-energy coherent hadron-nucleus scattering make a nucleus less
transparent. Important point to emphasize  is that  high-energy particles
interact with different nucleons of the nucleus at different moments of
time: $t_1~-~t_2=~c~(z_1-z_2)$ (that is  high-energy processes develop along
the light-cone). So the approximation of frozen configurations in the nucleus
used in the Glauber approximation seems, at first sight, questionable. However,
a theoretical analysis  based on the light-cone quantum mechanics of nuclei has
found \cite{FS81} that the conventional Glauber formulae can be safely used for
description of high-energy processes where contributing nucleon Fermi momenta
are not too large. In conventional quantum mechanics, the frozen nucleus
approximation emerges from the condition that intermediate nuclear states of
importance have excitation energies very small relative to the projectile
energy.
The application of this technique to the calculation of $(e,e'p)$ cross
sections
is straightforward  though it requires serious modifications due to the
different
collision geometry. There is no incident hadron, while a fast nucleon is
produced
in any point of the nucleus so that the  final expressions are rather different
from those in \cite{MNW,YN}. We will focus on the effects of nucleon
correlations
for the nuclear transparency in $A(e,e'N)$ reactions at high $Q^2$ as well as
on
the effects of the nuclear shells.

The  first question we address is  whether correlation effects considered in
hadron-nucleus scattering \cite{MNW,YN}  are relevant for the propagation of a
fast nucleon (produced in a hard scattering) through  nuclear matter. The
effects
of nucleon correlations for nuclear transparency in the $A(e,e'p)$ reaction
were
considered by several authors\cite{BFP92,LM,NICO93,KHO,RINAT,WALET} in
connection with the
recent experimental investigations of the color transparency. Benhar et al
\cite{BFP92} and  Lee and Miller \cite{LM} have suggested a modification of the
optical model approximation to include  the nucleon correlations. They have
found
that nucleon correlations may significantly (by $\sim 20\%$) increase the
$(e,e'p)$
cross section   and considerably suppress the onset of color transparency (CT)
effects especially at intermediate $Q^2$. However, in the papers
\cite{FSZ,NICO93,KHO}
the effects of correlations were estimated to be no more than $5\%$. We
demonstrate
here that the role of nucleon correlations  depends sensitively on the
experimental
kinematics. So, we will consider how the shell structure of nuclei observed at
medium energies (for a review, see Ref.\cite{MGFR}) should reveal itself in
high
$Q^2$ $A(e,e'N)$ reactions.

To analyze effects of nucleon correlations we consider here the limiting case
of
coherent final state interactions (FSI), when knocked-out protons rescatter off
the
residual nucleus coherently and the final state energy of the residual nucleus
is
fixed and known. A second case of incoherent rescattering, when  all elastic
rescatterings of the proton are allowed and the sum over the final states of
residual nucleus  is performed,  will be discussed elsewhere. The results
obtained
for coherent final state interactions  allow us to calculate nuclear
transparency at
a fixed value of the ejected  nucleon momentum and missing energy. Therefore
they
could be directly compared with current and planned experiments which have
limited
momentum and angular acceptance.

The paper is organized as following.

In section 2 we  present detailed calculation of the coherent FSI
on the basis of  the Glauber approximation. By  decomposing the
ground state wave function over the contribution of two-nucleon
correlations we derive the formulae for the description of
the  $(e,e'N)$  processes in the case of fixed missing energy
characterizing particular shells. The deduced formulae take into
account  the ground and  the final nuclear state correlations in
a selfconsistent way.

In section 3,  formulae  obtained  in  section 2  are used to calculate
the nuclear transparency for a proton knocked-out by the virtual photon.
Qualitative calculations using a uniform density model of the nucleus,
point-like approximation for $NN$ scattering amplitude and $\Theta$-function
type of $NN$-correlations allows us to obtain analytic results. Quantitative
analysis uses Hartree-Fock single-nucleon wave functions, realistic
parameterization of the $NN$ scattering amplitude and correlation effects
taken from the current calculations of the nuclear matter. The obtained
formulae are extended also to the case when electron produced a small size
"nucleon" wave packet which expands while propagating through the nucleus.
This gives the possibility of investigating how nucleon correlations influence
the onset of color coherent effects in  reactions with the excitation of
certain
shell levels. It is demonstrated that in some cases color transparency may even
lead to a decrease of the $(e,e'N)$ cross section with increase of $Q^2$.  This
effect has been considered in Ref. \cite{FSZ}, where, however, nucleon
correlations
were not included in the  theoretical analysis.

In section 4 we summarize  the basic results of the paper.

The Appendix contains necessary definitions and sum rules for the two nucleon
density function.

\section{Coherent Final State Interaction}

Within the nonrelativistic theory of nuclei and negleting antisymmetrization of
the
knocked-out and bound nucleons, the amplitude  of $A(e,e'N)$ scattering -
$F_{f,0}$
is given by the formulae:
\begin{eqnarray}
F_{f0}^{\gamma^*A} & = &
<\Psi_{\vec p_f}^{(-)},\Psi_f^{(A-1)}|\hat T_{\vec q}|\Psi_0^{(A)}>
\nonumber \\
 & = & \sum\limits_{j}\int d^3r_jd^3\{r_k\}
\Psi_{p_f}^{(-)}(\vec r_j)\Psi_f^{(A-1)}(\{\vec r_k\})
T^{em}(Q^2)
 e^{i\vec q\vec r_j}\Psi_0^{(A)}(\vec r_j,\{\vec r_k\}),
\label{F_gen}
\end{eqnarray}
where $\Psi_0^{(A)}$ is the ground state wave function of nucleus, and
$\hat T_{\vec q} \equiv T^{em}(Q^2)e^{i\vec q\vec r_j}$ is the one body
 electromagnetic current  operator.  In principle, $T^{em}$ should depend on
the presence of other nucleons due to off shell effects; however,
in this paper we restrict ourselves to the contribution of nucleons with
small Fermi momenta where off shell effects seem to be a small
correction. The position of knocked out nucleon $j$ is  $\vec{r}_{j}$;
$\vec q$ and $-Q^2$ are the three momentum and mass-square of the
virtual photon, $\Psi_{\vec p_f}^{(-)}$ is the  wave function
of knocked-out nucleon $j$; and   $\Psi_f^{(A-1)}$ is the wave function
of the residual $(A-1)$ nucleus. For simplicity we denote the
$\{r_k\}~\equiv~r_1,~..'..,~r_A$, where the coordinate of knocked-out
nucleon is excluded.

It is  well known from the low energy studies that the cross section of
exclusive $A(e,e'p)$ processes depends strongly  on the value of the missing
energy $E_m$ which characterizes the binding energy of the knocked-out
proton, as well as  the excitation energy of the residual $(A-1)$
nucleus.
  If $E_m$ is fixed and does not exceed the characteristic value for
the nuclear shell excitations $(\lesssim 50 MeV)$ (which is a natural
condition for the experiments searching for the color transparency
phenomenon) the final state interactions of the knocked-out nucleons
with the residual nucleus are dominated by their coherent rescattering
off the $(A-1)$-hole residual nucleus.

Neglecting the antisymmetrization between $r_j\leftrightarrow \{r_k\}$
the coordinate $r_j$   is  the knocked-out nucleon's coordinate.
Then within the  impulse approximation for the $\gamma^*N$ interaction
and Glauber approximation  for the interaction of the fast nucleon with
the rest of the nucleus, the amplitude $F_{f,0}$  of the process where
a nucleon is knocked-out from a specific orbit-$h$, leaving the
residual nucleus in a $(A-1)$-hole state, is expressed as follows:
\begin{eqnarray}
T_h^{\gamma^*A}\equiv F_{(A-1)-hole,0} = \int d^3r_1 d^3\{r_k\}
\Psi_0^{A}(r_1,\{r_k\})\cdot
\Psi_{h^{-1}}^{A-1^+}(\{r_k\})\cdot T^{em}(Q^2) e^{i\vec q\vec r_1}
\nonumber \\
\times e^{-i\vec p_i\vec r_1}\prod_{i=2}^{A}\left [1-\Gamma^N(b_1-b_i)
\cdot \Theta(z_i-z_1) \right ].
\label{T1}
\end{eqnarray}
Here   $\Psi^{A-1}_{h^{-1}}(\{r_k\})$ - is the wave function of
$A-1$-hole state of  residual nucleus and  $\vec p_i =\vec p_f-\vec q$.
To simplify the formulae, we shall always denote the coordinate of the
knocked-out nucleon as $r_1$ and will omit the sum over various nucleons $j$.
Here $\vec p_f$  is  the momentum of the knocked out nucleon.
The  rescatterings of the knocked out nucleon off the individual nucleons
of the residual nucleus are  described in eq. (2) by the product of functions
$\Gamma^{N}$. The profile function $\Gamma^N(b)$ is expressed through  the $NN$
amplitude ($f^{NN}$) as:
\begin{equation}
\Gamma^N(b) = {1\over 2\pi i k }\int \exp{(i{\vec k_t \vec b})}
\cdot f^{NN}({\vec k_t}) d^2k_t,
\label{gamma}
\end{equation}
where for  $f^{NN}$ we use the normalization
$Im f^{NN}~=~{k\over 4\pi}\sigma_{tot}e^{{b\over 2}t}$.

To calculate the amplitude  $T_h^{\gamma^*A}$ given by  eq.(\ref{T1})
it is convenient to approximate the ground state wave function according to
Ref.\cite{WS} as a product of the Slater determinant, representing the
uncorrelated ground state wave functions $\psi_{n}(r_{i})$ and Jastrow-type
correlated basis function $C_{h}(r_{i},r_{k})$:
\begin{eqnarray}
\Psi_0^{A}(r_1,...,r_A) & = & N(A!)^{-{1\over 2}} det|\psi_n(r_i)|
\prod\limits_{k>i=1}^{A}(1 + C_h(r_i,r_k))\nonumber \\
& = & N\sum\limits_{h} \omega_h
\phi_h(r_1)\prod_{k>1}(1 + C_h(r_1,r_k)) \Psi_{h^{-1}}^{(A-1)}(\{r_k\})
\label{cr0}
\end{eqnarray}
Here $w_{h}^2$ is the occupation probability for the nucleon orbital $h$.
In the last part of  eq.(\ref{cr0}) we modify the correlated basis
representation by introducing the single nucleon wave functions -
$\phi_h(\vec r_i)$, which represent the overlap integral between exact
$A$-body ground state wave function -$\Psi_0^{A}$ and $A-1$-body  wave
functions of the residual nucleus-$\Psi_{h^{-1}}^{(A-1)}(\{r_k\})$.
Single nucleon wave functions are normalized as follows:
\begin{equation}
\int|\phi_h(r)|^2d^3r = 1
\end{equation}
and \cite{MGFR,KNP}
\begin{equation}
\rho(r) = \sum\limits_{h}\omega_h^2|\phi_h(r)|^2,
\end{equation}
where $\rho(r)$ is the single nucleon density function defined according
to eqs.(\ref{a2}), (\ref{def2}), (\ref{n1}), which can be taken either
from nuclear many-body calculations or from experimental data.
Terms $C_h(r_1,r_k)$ in eq.(\ref{cr0}) parameterize ground state pair
nucleon correlations between detected nucleon $1$ and undetected nucleons -$k$.
The factor $N\approx 1+{\cal O}(l_c^3/R_A^3)$, accounts for the proper
normalization,  where $l_c$ - characterize the correlation lengths between
nucleons\cite{MN} and $R_A$ is the  nuclear radius (the  normalization factor
$N$ is further discussed  in the Appendix). In the case when three-nucleon
correlations in a nucleus can be  neglected $N\approx 1$.

Correlations enter eq.(\ref{cr0}) in two ways, through the correlation of
the struck nucleon with nearby nucleons (via $C_h(r_1,r_k)$ functions) and
through the correlations between "spectator" nucleons which are contained
in the function  $\Psi^{A-1}_{h^{-1}}$. Substituting, in eq.(\ref{T1}),
the nuclear ground state wave function by  eq.(\ref{cr0}) we obtain:
\begin{eqnarray}
T_h  & = & \sum\limits_{h'} \int d^3r_1 d^3\{r_k\}
\omega_{h'} \phi_{h'}(r_1)
\prod_{k=2}^{A}(1+C_{h'}(r_1,r_k))\Psi_{h'^{-1}}^{(A-1)}(\{r_k\})\cdot
T^{em}(Q^2)\cdot e^{-i\vec p_i\vec r_1}
\nonumber \\
& & \times \prod_{i=2}^{A}\left[1-\Gamma^N(b_1-b_i)\cdot\Theta(z_i-z_1)\right]
\Psi_{h^{-1}}^{(A-1)^+}(\{r_k\})
\nonumber \\ &  = &
\sum\limits_{h'} \int d^3r_1 d^3\{r_k\}
\omega_{h'}\phi_{h'}(r_1)\cdot T^{em}(Q^2)\cdot
e^{-i\vec p_i\vec r_1}
\prod_{i=2}^{A}\left [1-\Gamma^N(b_1-b_i) \cdot \Theta(z_i-z_1) \right ]
\nonumber \\
& & \times \left[\prod_{k=2}^{A}(1+C_{h'}(r_1,r_k))\right]\cdot
\rho_{h',h}^{A-1}(\{r_k\}).
\label{T_A1}
\end{eqnarray}
To simplify formulae, we introduce the $A-1$-body density matrix as:
\begin{equation}
\rho_{h'h}^{A-1}(\{r_k\}) \equiv  \Psi_{h'^{-1}}^{(A-1)}(\{r_k\})\times
\Psi_{h^{-1}}^{(A-1)^+}(\{r_k\}),
\label{rom0}
\end{equation}
which satisfy (as a consequence of the  orthogonality condition described in
Appendix) the following sum rules:
\begin{equation}
\int \rho_{h'h}^{A-1}(\{r_k\}) d^3\{r_k\} = \delta_{h',h},
\label{rom1}
\end{equation}
and
\begin{equation}
\rho^{A-1}(\{r_k\}) = \sum\limits_{h}\omega_h^2|\rho_{h,h}^{A-1}(\{r_k\})|^2,
\label{rom2}
\end{equation}
 where  $\rho^{A-1}(\{r_k\})$ is the conventional $A-1$-body density function.

In eq.(\ref{T_A1}) hard electromagnetic and soft hadron-nucleus scattering are
separated since we ignore the off-energy-shell effects in   $T^{e.m.}(Q^2)$.
To calculate nuclear effects we may use the methods developed for the
calculation
of  hadron-nucleus scattering \cite{MNW,MN,YN} provided the  $(A-1)$ nucleon
density
matrix (eq.({\ref{rom0})) of residual nucleus is known. However, this matrix is
practically unknown  now. So to evaluate transitions to certain nuclear levels,
we
will neglect the contribution of nondiagonal $h \rightarrow h'$ transitions
basing
on the following reasons:
i) our interest is in the processes where fixed missing energy is small
($\lesssim 50~MeV$) and we consider nuclei which have clear resolved
shell structure (e.g. $^{12}C$);
ii) the nondiagonal transitions from large $E_{m}$-states to small
$E_{m}$-state are strongly suppressed since FSI leads to further
increase of the overall missing energy;
iii) overlap integrals for the transition from small $E_m$-states to
large $E_m$-state are suppressed too  \cite{MGFR,KNP};
iv) the ability to fix missing momenta independently of the missing
energy in the considered kinematics, allow to  suppress further the
nondiagonal transitions, ( for example, in the case of $p_i\approx 0$
the contributions from $l\neq 0$ orbits are suppressed as compared to the
$l=0$ one);
v) nondiagonal $h \rightarrow h'$ transitions are a correction to the
contribution of nucleon correlations into nuclear transparency.
However we find below that effect of correlations in the processes we consider
is smal. Neglecting the nondiagonal transitions between the residual nuclear
states, we expand  the  $(A-1)$ body density function for a particular hole
-$h$, similar to Refs. \cite{MN,YN} through the (two, three, etc.)-body
correlation functions:
\begin{equation}
\rho_{h,h}^{A-1}(\{r_k\})   =
\rho(r_2)\times ... \rho(r_A) + \sum\limits_{i,j}g_{h}(r_i,r_j)
\cdot\rho(r_2)\times ...\rho(r_i)\times ..\rho(r_j) ..\rho(r_A)
\  \ + \  \ . . .  ,
\label{CON}
\end{equation}
where $\rho(r_i)$ is the above defined nucleon single density function.
In the approximation when only pair  nucleon correlations are kept,
it follows  from eqs.(\ref{rom1}) and (\ref{rom2}) (cf. Appendix) that:
\begin{equation}
\int g_{h}(r_i,r_j)\rho(r_i)\rho(r_j)d^3r_id^3r_j = 0
\label{rom3}
\end{equation}
Using this  decomposition of $(A-1)$-body density matrix (eq.(\ref{CON})),
we obtain for FSI practically the same functional form as that obtained
in \cite{MN,YN} for $hA$ scattering within the Glauber approximation,
when two body correlations are taken into account.  The major difference
from \cite{MN,YN} is the different geometry of $eA$ collisions as compared
to $hA$ scattering.  This difference is accounted for in eq.(\ref{T_A1}) in
the limits of integration over the coordinate $z_{1}$   of the knocked out
nucleon.
Taking into account the normalization conditions for the correlation function
$C(r_i,r_j)$ (see Appendix) and $g_{h}(r_i,r_j)$ (eq.(\ref{rom3})) we obtain
for the amplitude $T_h$  the expression similar to that in Ref.\cite{MN}:
\begin{eqnarray}
 T_h^{\gamma^*A}  & = &  \int d^3r_1 \cdot\omega_h\phi_h(r_1)
\cdot T^{em}(Q^2) \cdot e^{-i\vec p_i\vec r_1}
\nonumber \\
& \times &  \left[1-\int\limits_{z_1}d^3r
\left(1+C_{h}(r_1,r)\right)\rho(r)\Gamma(b_1-b)\right]^{(A-1)}
\cdot {\cal P}_{A-1}((1+C_h),\Gamma),
\label{T_BE}
\end{eqnarray}
where the factor  ${\cal P}_{A-1}((1+C_h),\Gamma)$, characterizes influence
of correlations between undetected nucleons on FSI of knocked-out
nucleon\cite{MN}:
\begin{eqnarray}
& & {\cal P}_{A-1}((1+C_h),\Gamma)  =
\sum\limits_{m=0}^{{A-1\over 2} or {A-2\over 2}} {(A-1)!\over (A-1-2m)!m!}
\nonumber \\
& & \times   \left[{{1\over 2}\int\limits_{z_{2,3}>z_1}d^3r_2d^3r_3
\left(1+C_{h}(r_1,r)\right)g_{h}(r_2,r_3)\rho(r_2)\rho(r_3)
\Gamma(b_1-b_2)\Gamma(b_1-b_3)
\over (1-\int\limits_{z>z_1}d^3r\left(1+C_h(r_1,r)\right)\rho(r)
\Gamma(b_1-b))^2}\right]^m.
\nonumber \\
\label{P_corr}
\end{eqnarray}
Eqs.(\ref{T_BE}) and (\ref{P_corr}) show that correlations between undetected
nucleons enter similarly to $hA$ scattering \cite{MN,YN}, while correlations
between the knocked-out nucleon and undetected nucleons enter via the rescaling
of the single nucleon density function by the factor $(1+C_h(r_1,r))$.

At large $A$,  eqs.(\ref{T_BE}), (\ref{P_corr})  can be considerably simplified
by keeping in the factor $[. \ .]^m$  in eq. (\ref{P_corr}) only the terms
which
grow with $A$. The formulae obtained for large $A$  resemble the  optical limit
of Glauber approximation formulae:
\begin{equation}
T_h^{\gamma^*A} = \int d^3r_1 \cdot \omega_h\phi_h(r_1)\cdot T^{em}(Q^2)\cdot
e^{-i\vec p_i\vec r_1} e^{-\int\limits_{z_1}\Gamma(b_1-b)
\tilde n(r) d^3r},
\label{T_coh}
\end{equation}
where $\tilde n(r)\equiv (A-1)\cdot\tilde \rho(r)$. The modified nuclear
density is:
\begin{eqnarray}
\tilde\rho(z,b) & = &   \left[1 + C_{h}(r_1,r)\right]\left( 1
- {A-1\over 2}\int\limits_{z_1}\Gamma(b_1-b')g_{h}(r,r')1
\rho(r')d^3r'\right)\rho(r)
\nonumber \\
& \approx & \rho(r)\left[ 1 + C_{h}(r_1,r)
- {A-1\over 2}\int\limits_{z_1}   \Gamma(b_1-b')
g_{h}(r,r')\rho(r')d^3r'\right],
\label{tlrho}
\end{eqnarray}
where, at the last step, we neglect the term proportional to the square of
correlations, since its contribution is comparable to higher order correlations
which were neglected  earlier. In practice eq.(\ref{T_coh}) is applicable
starting
from $A\ge 4 $, with accuracy comparable to the accuracy of exponentiation of
$(A-1)$-power function in eq.(\ref{T_BE}), estimated as
$\sim{\cal O}({1\over 2A}[\int\limits_{z>z_1}\Gamma(b_1-b)\tilde n d^3r]^2)$.

In the limit when correlations are neglected the derived equations coincide
with the
formulae of the Glauber approximation in the independent particle approximation
for
the nuclear wave function. When FSI is neglected,  the derived formulae lead to
plane
wave impulse approximations within the generalized shell model, where
correlations
in the ground state wave functions are taken into account (see e.g.
\cite{MGFR,DIEP,VSTN}).

The interesting feature of eqs.(\ref{T_coh}), (\ref{tlrho}) is that account of
nucleon
correlations influence the final state interaction in two opposite ways. Due to
the
second term in eq.(\ref{tlrho}), correlations lead  to decrease of the
effective
nuclear densities ($C_{h}(r,r')~<~0$ ) and therefore to increase of the
transparency of
nuclear matter for the knocked-out nucleon. This effect reflects the  presence
of a hole
around the scattered nucleon in the ground state wave functiosendn. This effect
has been
previously  mentioned  in Ref.\cite{FSZH90} and analyzed at length in
\cite{BFP92},
within the optical approximation and for cross sections integrated over missing
momentum
and energy.

However, the contribution of correlations in the third term of eq.(\ref{tlrho})
leads to increase of the effective nuclear density and as a result, nuclear
matter become more opaque for the knocked-out nucleon. Same  effect  was found
in
Refs.\cite{MNW,MN,YN} for  high energy hadron-nucleus scattering.
Similar effect was discussed also in Refs.\cite{NICO93,RINAT,WALET} for the
cross section
of $(e,e'p)$ reaction summed over the final states of residual nucleus and
integrated over
the proton momentum. In this case this  contribution into the overall
correlation effect
is practically negligible \cite{RINAT,WALET}.

To calculate the cross section of the semiexclusive  $(e,e'N)$ reaction we
use the distorted wave impulse approximation (DWIA) where the  cross section
can
be represented as follows:
\begin{equation}
{d^6\sigma \over d\epsilon_2d\Omega_2d^3\vec p_f} =
\sigma_{eN} \cdot S_A(\vec p_i,Em,\vec p_f),
\end{equation}
where $\sigma_{eN}$ is proportional to the cross section of electron scattering
off a
bound nucleon. We restrict ourselves to  the case of  reactions where nucleon
Fermi motion
is small (i.e., small missing momenta). For large missing momenta, more
accurate treatment of
multistep processes and relativistic effects is necessary.

If the specific shell is fixed the DWIA spectral function  can be written as
\cite{MGFR}:
\begin{equation}
S_A(\vec p_i,E_m,\vec p_f) =
n_h(E_m)\cdot\mid\Phi_h(\vec p_i,\vec
p_f)\mid^2,
\end{equation}
where  $n_h(E_m)$ characterizes the strength of the
shell and proportional to the shell occupation probabilities $\sim \omega_h^2$.
$\Phi_h(\vec p_i,\vec p_f)$ is the distorted momentum distribution of the
nucleons
for $h$-shell. Using  eq.(\ref{T_coh}) we  obtain for
$\mid\Phi_h(\vec p_i,\vec p_f)\mid^2$:
\begin{eqnarray}
\mid \Phi_h(\vec p_i,\vec p_f)\mid^2 =
\left| \int d^3r_1\Psi_h(r_1) e^{-i\vec p_i\vec r_1}
\times\left[e^{-\int\limits_{z>z_1}\Gamma(b_1-b)\tilde
n(r)d^3r} \right ]\right| ^2.
\label{PHI_CH}
\end{eqnarray}

\section{Nuclear transparency}

We  use   eq.(\ref{T_coh}) for the scattering amplitude and eq.(\ref{PHI_CH}),
for the distorted momentum distributions to calculate  $(e,e'N)$ scattering on
nuclei.

In this section we will consider the effects which are due to the final state
interaction of knocked - out nucleons. The convenient quantity to characterize
the
FSI (see e.g. Refs.\cite{NE18,BFP92,FSZ,FFLS}) is the ratio of the measured
cross
section of the $(e,e'N)$ reactions and the  cross section calculated within the
plane
wave impulse approximation (PWIA).  In the case of complete nuclear
transparency
this ratio will  equal   unity. The corresponding  theoretical quantity is the
ratio
of the cross sections  calculated  with and without FSI.

In the theoretical analysis  a convenient quantity is the  transparency
corresponding
to a transition when  the particular nuclear shell is   fixed (no summation
over the
final states of residual nucleus):

\begin{equation}
T_r^h \equiv \left( {\sigma^{EXP}\over
\sigma^{PWIA}}\right)_h =
{\mid \Phi_h^{DWIA}(p_i,p_f)\mid^2 \over
\mid \Phi_h^{PWIA}(p_i)\mid^2 },
\label{Tr_SH}
\end{equation}
where  $\mid \Phi_h^{DWIA}(p_i,p_f)\mid^2$ is given by eq.(\ref{PHI_CH}).

\subsection{Qualitative estimates }

To visualize the  role of nucleon-nucleon correlations it is worth considering
first
a highly simplified  model of uniform  nuclear density. We will also treat
nucleons as
point-like and furthermore approximate correlation functions as
$\Theta$-functions:
\begin{equation}
C_h(x) \approx g(x) \approx -\Theta(l_c - x),
\label{TH}
\end{equation}
where $l_c$ is the  correlation length defined as\cite{MN}:
\begin{equation}
l_c = -\int_0^\infty g(r) dr,
\label{LC}
\end{equation}
Here $g(r)$ is the  correlation function calculated  within the realistic
theory of
nucleus. We will  omit  $T^{em}(Q^2)=1$  in the next following analysis since
electromagnetic form factors of a nucleon are canceled in  eq.(\ref{Tr_SH}) for
nuclear
transparency provided the off-shell effects are neglected.

The profile function of point-like nucleon is  (cf eq.(\ref{gamma})):
\begin{equation}
\Gamma(b_1-b_2) = {\sigma_{tot}\over 2}\cdot \delta^2(b_1-b_2).
\label{PL}
\end{equation}

Using the  above approximations and assuming that,
\begin{equation}
\int_{z_1}g(z-z_1)\rho(z,b_1)dz \approx - 2l_c\rho(z_1,b_1)
\label{ASSUM}
\end{equation}
we obtain  for the exponent in   eq.(\ref{T_coh}) :
\begin{equation}
-\int\limits_{z_1}\Gamma(b_1-b)\tilde n(r) d^3r =
-{\sqrt{R^2-b_1^2} - z\over 2\lambda}
- {l_c\over \lambda}\left[{\sqrt{R^2-b_1^2}
- z\over4\lambda} - 1\right],
\label{CH_coh}
\end{equation}
where $R$-is the nuclear radius and $\lambda = {1\over \sigma\rho_0}$ is the
mean
free path, and $\rho_0$ is the uniform  density. Approximating  the one-body
wave function
as uniform:
\begin{equation}
\phi_h = {\Theta(|R-r|)\over ({4\pi\over 3}R^3)^{1\over 2}}
\label{phi_h}
\end{equation}
we calculate the  amplitude $F^{f0}$ with coherent FSI in  the case of
$\vec p_i = \vec p_f - \vec q = 0$ for transition to the (A-1) nucleon system:
\begin{equation}
F^{coh} = {e^{{L\over \Lambda}} ({4\pi\over 3}R)^{1\over 2}\over
{1\over \Lambda} + {L\over 2\Lambda^2}}\cdot
{3\over 2}\left [1 +{2\left(e^{-({1\over \Lambda}+{L\over 2\Lambda^2})}
(1 + {1\over \Lambda} + {L\over 2\Lambda^2}) - 1\right)\over
({1\over \Lambda}+{L\over 2\Lambda^2})^2}\right]
\label{F_f0}
\end{equation}
and for  the transparency defined as in eq.(\ref{Tr_SH}) (for certainty we
consider
the case k=0):
\begin{equation}
T_r^{coh} = {|F^{coh}|^2\over |\phi_h(k=0)|^2} =
{e^{{2L\over \Lambda}}\over
({1\over \Lambda} + {L\over 2\Lambda^2})^2}\cdot
{9\over 4}\left [1 +{2\left(e^{-({1\over \Lambda}+{L\over 2\Lambda^2})}
(1 + {1\over \Lambda} + {L\over 2\Lambda^2}) - 1\right)\over
({1\over \Lambda}+{L\over 2\Lambda^2})^2}\right]^2,
\label{Tr_f0}
\end{equation}
where $L\equiv {l_c\over R}$ and $\Lambda\equiv  {\lambda\over R}$.

To visualize the  effects of  correlations  in eqs.(\ref{Tr_f0})
it is convenient to normalize transparency to the corresponding
transparencies within the uncorrelated Glauber approximation and to
consider the  case when $A$  is sufficiently large and
${1\over \Lambda} \gg 1$, ${L\over \Lambda}={l_c \over \lambda} \ll 1$.

We obtain for this ratio:
\begin{equation}
\left({T_r\over T_r^{noncorr}}\right)^{coh}\approx
{e^{{2L\over\Lambda}}\over(1+{L\over 2\Lambda})^2} \approx
(1+{L\over  \Lambda}).
\label{COH}
\end{equation}

Eq.(\ref{COH}) clearly demonstrates that correlation between
detected nucleon and  non  detected nucleons (numerator
in eq.(\ref{COH})) and correlations between undetected
nucleons (denominator in eq.(\ref{COH})) enter differently into
nuclear transparency and the first effect  dominates.

The next important feature of coherent rescatterings for transitions
to a ground state is the strongly different $A$ dependence of nuclear
transparency compared to the case of incoherent final state interactions
$\sim A^{-{1\over 3}}$( see e.g. \cite{FFLS}). From eq.(\ref{Tr_f0}) we obtain:
\begin{equation}
T_r^{coh} \sim  \Lambda^2 \sim R^{-2} \sim  A^{-{2\over 3}}.
\label{A_COH}
\end{equation}
This considerably  oversimplified model of the nucleus demonstrates the
qualitative difference between coherent and incoherent final state
interactions.
The obtained formulae  show that fixing the final states of residual nucleus is
more
promising  for searching color transparency effects in nuclei  since CT effects
are
larger in this case.

\subsection{Quantitative calculations }

We present here numerical results for the case of $(e.e'p)$ scattering
off  $^{12}C$ and  use the kinematics where momentum of knocked-out proton is
equal to the transferred momentum: $\vec p_f = \vec q$. The distinguishable
shell
structure of $^{12}C$ allows us to outline the effects of shell structure on
the
nuclear transparency. In calculations of the $^{12}C$ ground state wave
functions
we use the Skyrme - Hartree - Fock model with correlated interaction
\cite{KNP}.

To describe the correlation properties of the nuclear ground state and final
$(A-1)-hole$ state, we assume that NN pair correlations are state independent.
This assumption is inferred from both  theoretical and experimental
observations
(see e.g. \cite{FS81,BERT,MCH}) indicating that the nuclear high momentum
components
(controlled mainly by short range NN correlations) are practically the same for
all
nuclei. This approximation is not reliable for the long range correlations,
where
strong density dependence (observed in Ref.\cite{PAN}) should be taken into
account.
For our calculations we use the correlation function $g(r)$ from the
calculation of
\cite{PAN,P} for standard nuclear density ( $ = 0.16~fm^{-3}$).\footnote
{ The function $g(r)$ used in present work is related  to the correlation
function
$g_0(r)$ obtained  in \cite{PAN,P} as $g(r) = {A\over A-1} g_0(r) - 1$.}
The  accuracy of such approximation depends on the overall size of correlation
effects
and use of the exact density-dependent correlation function would clearly
improve the
present calculations.

The profile functions in eq.(\ref{gamma}) have been calculated  using the
relation
\begin{equation}
\mid f(k_t)\mid^2 = {k^2\over \pi} {d\sigma\over dt}.
\end{equation}
For ${d\sigma\over dt}$ we use the phenomenological
 parameterization:
\begin{equation}
{d\sigma\over dt} = {\sigma_{tot}^2 \over 16\pi} (1 +
\alpha^2)exp(bt),
\end{equation}
where $\alpha = Re f / Im f$ and all parameters are taken from
\cite{DBR,SILV,PDB}.

In  fig.1 we present the results of calculations of the
$Q^2$ dependence of the nuclear transparency based on
eq.(\ref{Tr_SH}) for the proton knocked out from the $s$ shell.
We observe that two opposite effects of nucleon correlations reduce
the overall effect of correlations to the level of few percent.

To see the interplay of  the above discussed effects  with the
anticipated effects of color transparency (CT) we use
the quantum diffusion model (QDM) \cite{FFLS} to account
for the reduction of FSI for knocked-out proton due to CT
effects. For this purpose  we introduce the modified profile
function in eq.(\ref{gamma}) with the modified $NN$ scattering
amplitude \cite{double}:
\begin{equation}
f^{NN}(k_t,Q^2,l) \approx i{k\over
4\pi}\sigma_{tot}(l,Q^{2})
\cdot e^{{b\over 2 }t}\cdot
{G_{N}(t\cdot\sigma_{tot}(l,Q^{2})/\sigma_{tot})
\over G_{N}(t) },
\label{F_NN_CT}
\end{equation}
where $b$ is the slope of elastic $NN$-cross section  and
$G_{N}(t)$ ($\approx (1-t/0.71)^{2}$) is the Sachs form factor.
The last factor in eq.~(\ref{F_NN_CT}) accounts for the
difference between  form-factors  for point-like and average
configurations, which is estimated based on the observation that the $t$
dependence of ${d\sigma^{h+N\rightarrow h+N}/dt\sim~G_{h}^{2}(t)\cdot
G_{N}^{2}(t)}$. The effective $NN$ total cross section we calculate
using the (QDM) predictions \cite{FFLS}:
\begin{equation}
\sigma_{tot}(l,Q^2) = \sigma_{tot}\left\{\left( {l\over l_h}
+ {<r_t(Q^2)^2>\over <r_t^2>}(1-{l\over l_h})\right)\Theta(l_h-l) +
\Theta(l-l_h)\right\},
\label{sigma}
\end{equation}
where  $l_h = 2p_f/\Delta M^2$, with $\Delta M^2 = 0.7~GeV^2$.
${<r_t(Q^2)^2>\over <r_t^2>} \approx {1 GeV^2\over Q^2}$
- is the average transverse size squared of the configuration
produced in the interaction point.

In fig.2 we present the $Q^2$ dependence of the color transparency
effect in the kinematics of fig.1  (curves labeled "s-shell").
Fig.2 shows that the correlation effect becomes smaller  at high $Q^2$
since  according to Eqs.(\ref{T_coh})-(\ref{tlrho}) the reduced value of
the  cross section of $NN$ scattering  reduces sensitivity to
the changes of density function $\rho \rightarrow \tilde \rho$.
However correlations slightly  reduce the onset of CT effect
since the  nucleus becomes more transparent.

Another important effect, in the kinematics of coherent FSI, is different
manifestation  of  color transparency  for fixed $s$ and $p$ shells.  It
follows from fig.3 that in  the case of the proton knock-out from $p$ and $s$
shells in the kinematics where $\vec p_i = \vec p_f - \vec q \approx 0$  the
decrease of  FSI leads to    opposite effects on the  cross section for
$(e,e'p)$
reaction for s- and p-shells. In fig.4 the expected CT effects  calculated
using
eqs.(\ref{F_NN_CT},\ref{sigma}) normalized to the corresponding transparencies
within the Glauber approximation.

\section{Conclusions}

We investigated  the final state interaction of knocked out nucleons in high
$Q^2$ $(e,e'N)$ processes within the Glauber approximation, taking into account
the nucleon correlations in a consistent way in the nuclear ground state and
fixed $A-1$ final state wave function. The main conclusion is that nucleon
correlations affect  the nuclear transparency in two different ways:
correlations among  undetected nucleons make a nucleus less transparent, while
the correlations  among the detected  nucleon and undetected nucleons make the
nucleus more transparent.

The consideration  of nuclear shell structure shows that effects of
correlations
are on the level of few percent for coherent final state interaction.

By including color coherent effects we elaborate the observation of
Ref.\cite{FSZ}
that color transparency have different implications for the  excitation of
different
nuclear  shells. We conclude that $(e,e'p)$ reactions are more sensitive to the
color
coherent effects provided the particular nuclear shell is fixed.

\acknowledgments
This work was supported in part by the U.S. Department of Energy under
grants   DE-FG02-93ER40771 and DE-FG02-94ER40823, and Israel- USA Binational
Science Foundation  Grant No. 9200126. EJM is grateful for support from the
Alexander
von Humboldt Stiftung.

\section*{Appendix}


To construct the wave function of the nuclear ground state via
single-nucleon wave function - $\phi_h(r_1)$  and $(A-1)-hole$
wave function - $\Psi_{h^{-1}}^{(A-1)}(\{r_k\})$ we introduce
the $h$-state dependent correlation functions between nucleon at
$r_1$ and $A-1$ nucleons belonging to the rest of the nucleus:
\begin{equation}
 \Psi_0^{A}(r_1,...,r_A) = N\sum\limits_{h} \omega_h
\phi_h(r_1)\prod_{k>1}(1 + C_h(r_1,r_k)) \Psi_{h^{-1}}^{(A-1)}(\{r_k\})
\label{a1}
\end{equation}
where $\{r_k\}\equiv r_2, .... r_A$.
\begin{equation}
 \int |\Psi^A_0(\{r_k\})|^2 d^3 \{r_k\} =1
\label{a11}
\end{equation}
We choose normalization of  $\Psi_{h^{-1}}^{(A-1)}$ as
\begin{equation}
\int \Psi_{h^{-1}}^{(A-1)} \Psi_{h'^{-1}}^{(A-1)} d^3 \{r_k\} =\delta_{hh'}
\label{a12}
\end{equation}
The overall normalization factor $N\approx 1$ provided three(or more)-nucleon
correlations are neglected.
  To reproduce the formulae of the shell model we choose $\rho (r_1)$ as
\begin{equation}
\rho(r_1) = \int \rho_A(r_1,\{r_k\})d^3\{r_k\} =
\int |\Psi_0^{A}(r_1,\{r_k\})|^2d^3\{r_k\} = \sum\limits_h\omega_h^2
\phi_h^2(r_1),
\label{a2}
\end{equation}
where $\rho(r_1)$-is the single nucleon density function.

As a consequence of the orthogonality of $h$-state wave functions - eq.(36)
and eq.(\ref{a2}) the correlation functions $C_h(r_1,r_k)$ should obey the
following relations:
 \begin{equation}
\int\left[  \prod_{k>1}(1+C_h(r_1,r_k)) \prod_{k>1}( 1+
C_{h'}^+(r_1,r_k))-1\right]
\Psi_{h^{-1}}^{(A-1)}(\{r_k\})\Psi_{h'^{-1}}^{(A-1)^+}(\{r_k\})
d^3 \{r_k\} = 0.
\label{a3}
\end{equation}
if three  ...nucleon correlations are neglected. For $h=h'$ , within the
accuracy ${\cal O}(C_h^2)$ we obtain:
\begin{equation}
\int C_h(r_k,r_k)|\Psi_{h^{-1}}^{(A-1)}(\{r_k\})|^2 = 0.
\label{a4}
\end{equation}
For the practical aims  we neglect the   dependence
of the $(A-1)$-nucleus wave function
$|\Psi_{h^{-1}}^{(A-1)}(\{r_k\})|^2 \equiv \rho^{(A-1)}(\{r_k\})$
and the correlation function $C_h(r_1,r_k)\equiv C(r_1,r_k)$ on the
nucleon orbital h expressing them
through the correlation function $g(r_i,r_j)$ defined as:
\begin{eqnarray}
\int|\Psi_0^A(r_1,r_2..,r_A)|^2d^3r_1 d^3r_{i-1} d^3r_{i+1}
d^3t_{j-1}d^3r_{j+1}
d^3r_A & = &  \rho_2(r_i,r_j) = \nonumber \\
\left( (1 + g(r_i,r_j) \right)\cdot\rho(r_i)\cdot\rho(j),
\label{def}
\end{eqnarray}
where two-nucleon density function $\rho_2(r_i,r_j)$ is
the probability to  find $(i,j)$ nucleons $(i,j)$
in nucleus  simultaneously at the points  $r_i$ and $r_j$.
$\rho_2$ is normalized as:
\begin{equation}
\int\rho_2(r_i,r_j)d^3r_id^3r_j = 1.
\label{n2}
\end{equation}
 The  single nucleon density $\rho(r)$ is:
\begin{equation}
\int\rho_2(r,r')d^3r' = \rho(r).
\label{def2}
\end{equation}
It is  normalized  as:
\begin{equation}
\int\rho(r)d^3r = 1
\label{n1}
\end{equation}

Inserting  eq.(\ref{def}) to    eq.(\ref{n2}) and using eq.(\ref{n1})
we obtain the following sum rule  for the correlation function $g(r_i,r_j)$ :
\begin{equation}
\int g(r_i,r_j)\rho(r_i)\rho(r_j)d^3r_id^3r_j = 0
\label{ag1}
\end{equation}

Inserting   eq.(\ref{def}) to    eq.(\ref{def2})we obtain also:
\begin{equation}
\int g(r_i,r_j)\rho(r_i)d^3dr_i = \int g(r_i,r_j)\rho(r_j)d^3r_j = 0
\label{ag2}
\end{equation}

Expanding   the $A$ body density function through the above defined correlation
function $g$ as:
\begin{equation}
\rho^{(A)}(r_1,\{r_k\}) = \rho(r_1)\prod\limits_{k=2}^A
(1 + g(r_1,r_k))\rho^{(A-1)}(\{r_k\}),
\end{equation}
and comparing it with $A$ body density function defined by square
modulus of eq.(\ref{a1}) one obtain following relations between two
correlation functions:
\begin{equation}
|1 + C(r_1,r_k)|^2 = 1 + g(r_1,r_k)
\label{a8}
\end{equation}

Obtained   relation allows to  estimate the accuracy of the approximation
$N=1$. Using the eqs.(\ref{a1}) and (\ref{a8}) and the normalization condition
for
$\Psi_0^{A}(r_1,...,r_A)$ we obtain:
\begin{eqnarray}
1 & = & N^2\left[ 1 + \sum\limits_{i,j}\int
g(r_i,r_j)\rho(r_i)\rho(r_j)d^3r_id^3r_j \right.
\nonumber \\
& & + \sum\limits_{i\leftrightarrow j,k}\int g(r_i,r_j)g(r_i,r_k)
\rho(r_i)\rho(r_j)\rho(r_k) d^3r_id^3r_jd^3r_k
\nonumber \\
& & +  \sum\limits_{i\neq j\neq k \neq m}\int g(r_i,r_j)g(r_k,r_m)
\rho(r_i)\rho(r_j)\rho(r_k)\rho(r_m) d^3r_id^3r_jd^3r_k d^3r_m
\nonumber \\
& & +  \left.\sum\limits_{i\leftrightarrow j,k}\int
g(r_i,r_j)g(r_i,r_k)g(r_j,r_k)
\rho(r_i)\rho(r_j)\rho(r_k) d^3r_id^3r_jd^3r_k + . . .\right]
\label{a9}
\end{eqnarray}
Taking into account the   sum rules for function $g(r_i,r_j)$
(eqs.(\ref{ag1}) and (\ref{ag2})) we find that first nonvanishing term in
eq.(\ref{a9}) is proportional to $\sim g^3$. In the framework of the uniform
density   model of nucleus ( see section 4.1) we obtain that:
\begin{equation}
N\approx 1 + {\cal O}({l_c^3\over R_A^3})
\end{equation}
where $R_A$ is the nuclear radius and $l_c$ - is the correlation length
defined in eq.(\ref{LC}). Using the estimation $l_c\approx 0.74~fm$
Ref.\cite{MN}
for   nuclei with $A\geq 12$  the accuracy of condition   $N=1$ is better than
$1-2\%$.
The effect of neglected three-nucleon correlations is expected to be on same
level,
since they proportional to $\sim l_c^3$.


\newpage
\centerline{\bf Figure Caption}

\begin{itemize}

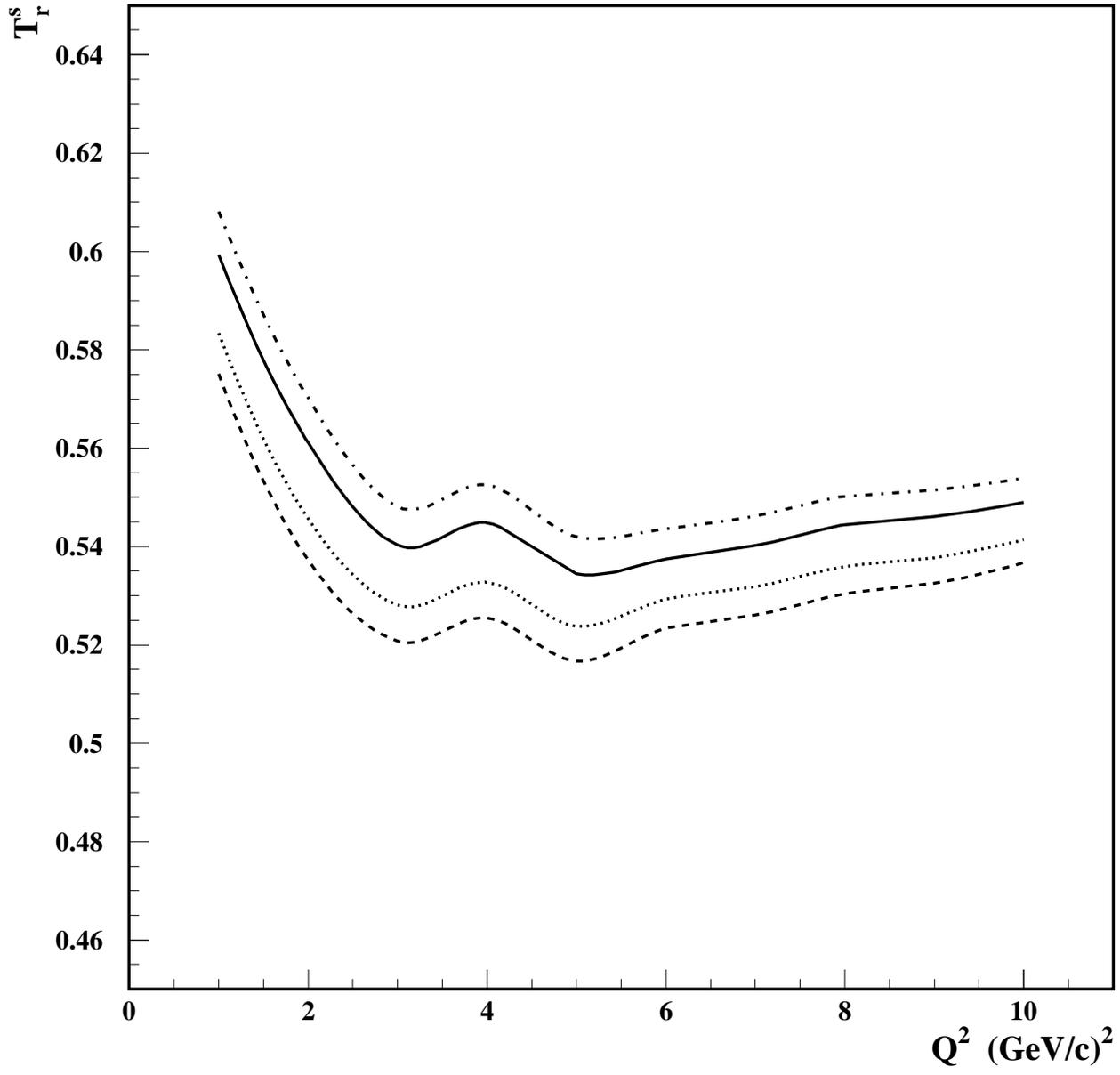
\item [ Figure 1.] The $Q^2$ dependence of the nuclear transparency $T_r$ of
$^{12}C$, calculated according to eq.(\ref{Tr_SH}) for the reaction of proton
knock-out from $s$-shell with $\vec p_i~=~0$ including the
energy dependence of the $NN $ amplitudes.
Dotted line - is the calculation
without correlation effects, dashed line - with the effects of correlation
between undetected nucleons, dash-dotted line - with the effects of correlation
between knocked-out proton and  undetected nucleons and solid line - with
overall correlation effects.

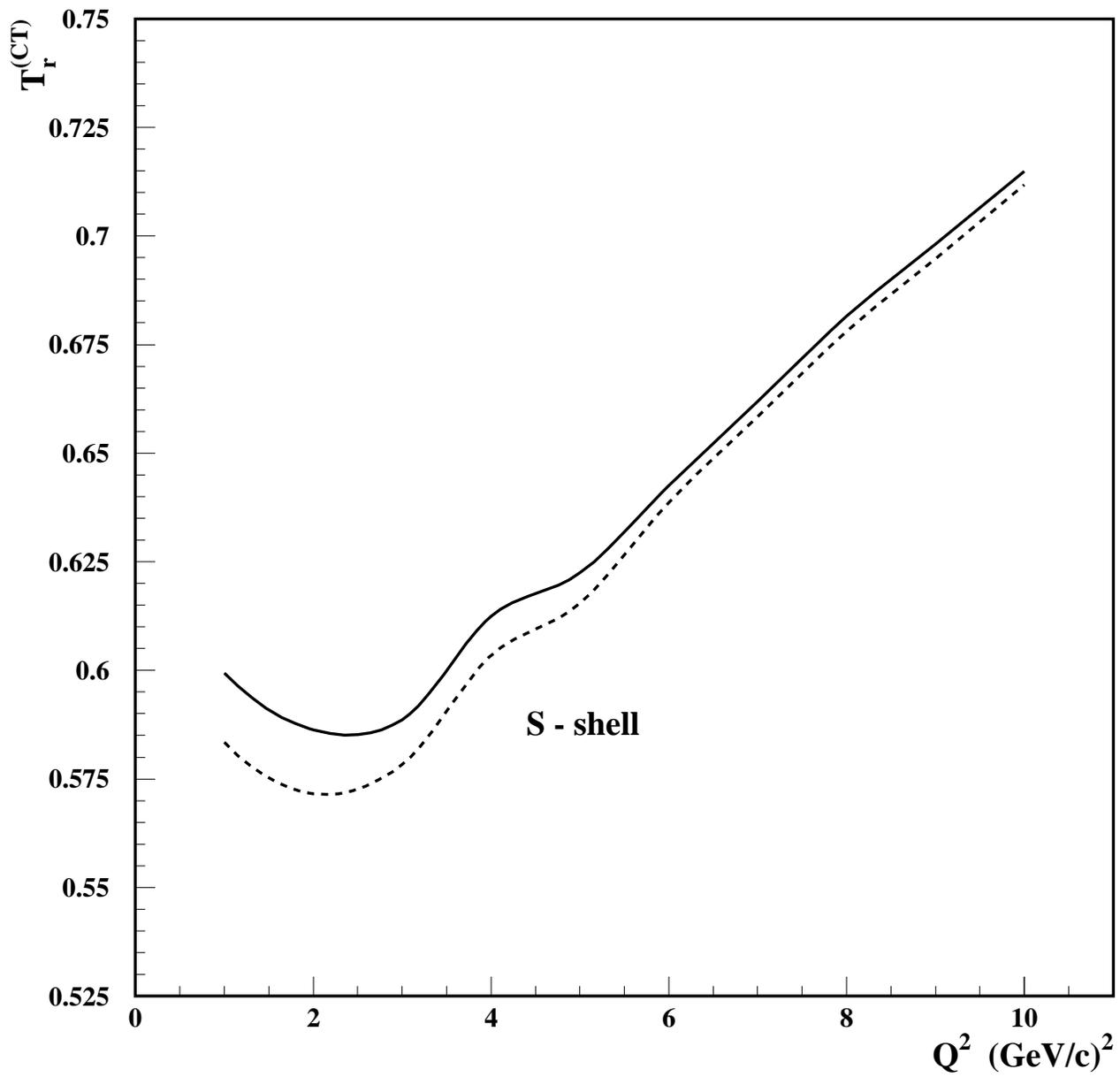
\item [ Figure 2.]  The $Q^2$ dependence of color transparency effect
defined by eqs.(\ref{Tr_SH}), corresponding to the kinematics of fig.1.
Dashed line - without,  solid line - with overall
correlation effects.

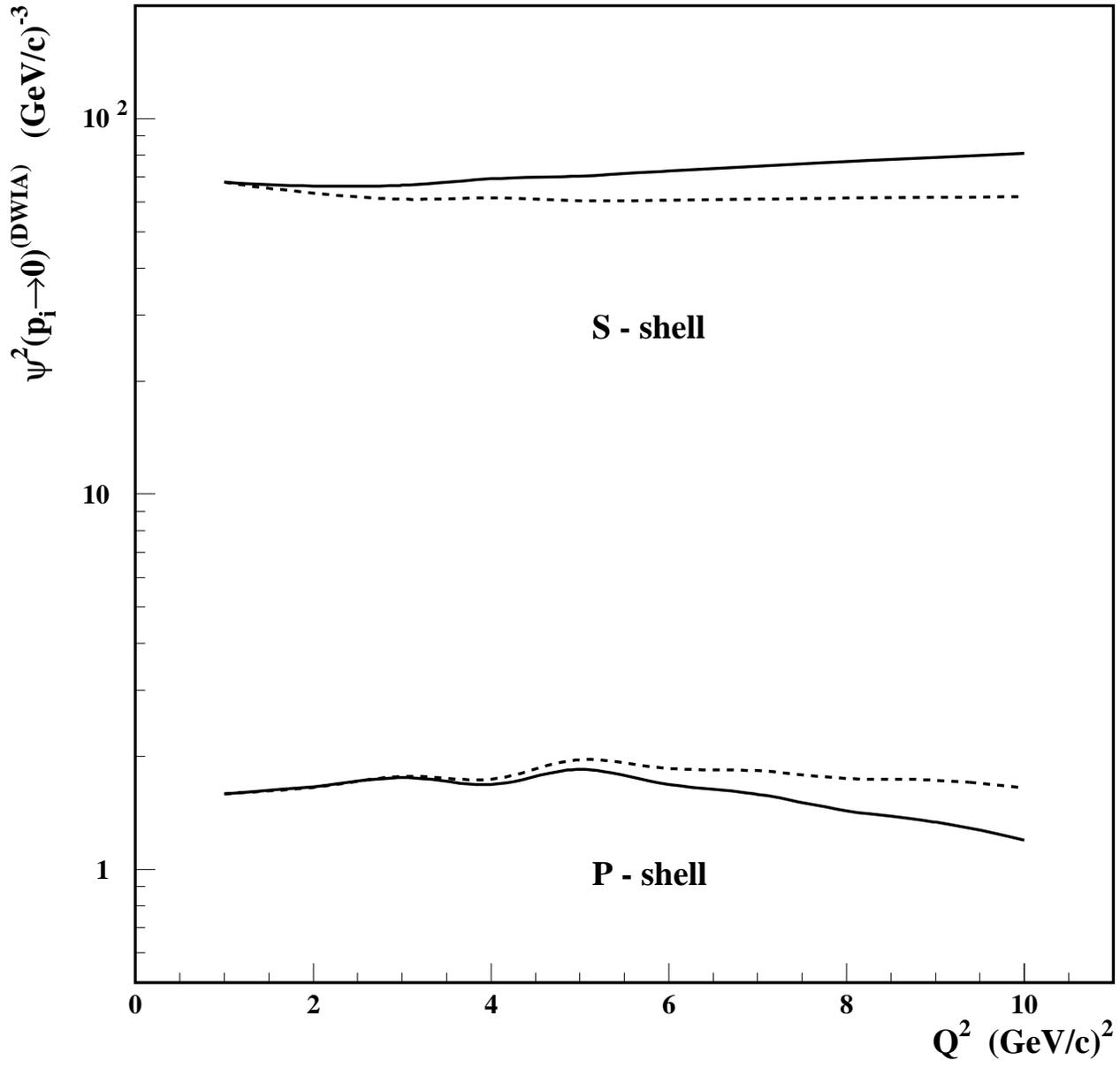
\item [ Figure 3.]  The $Q^2$ dependence of color transparency effect
in distorted momentum distribution of proton on $s$ and $p$ shells.
Dashed line - without, solid line - with overall correlation effects.

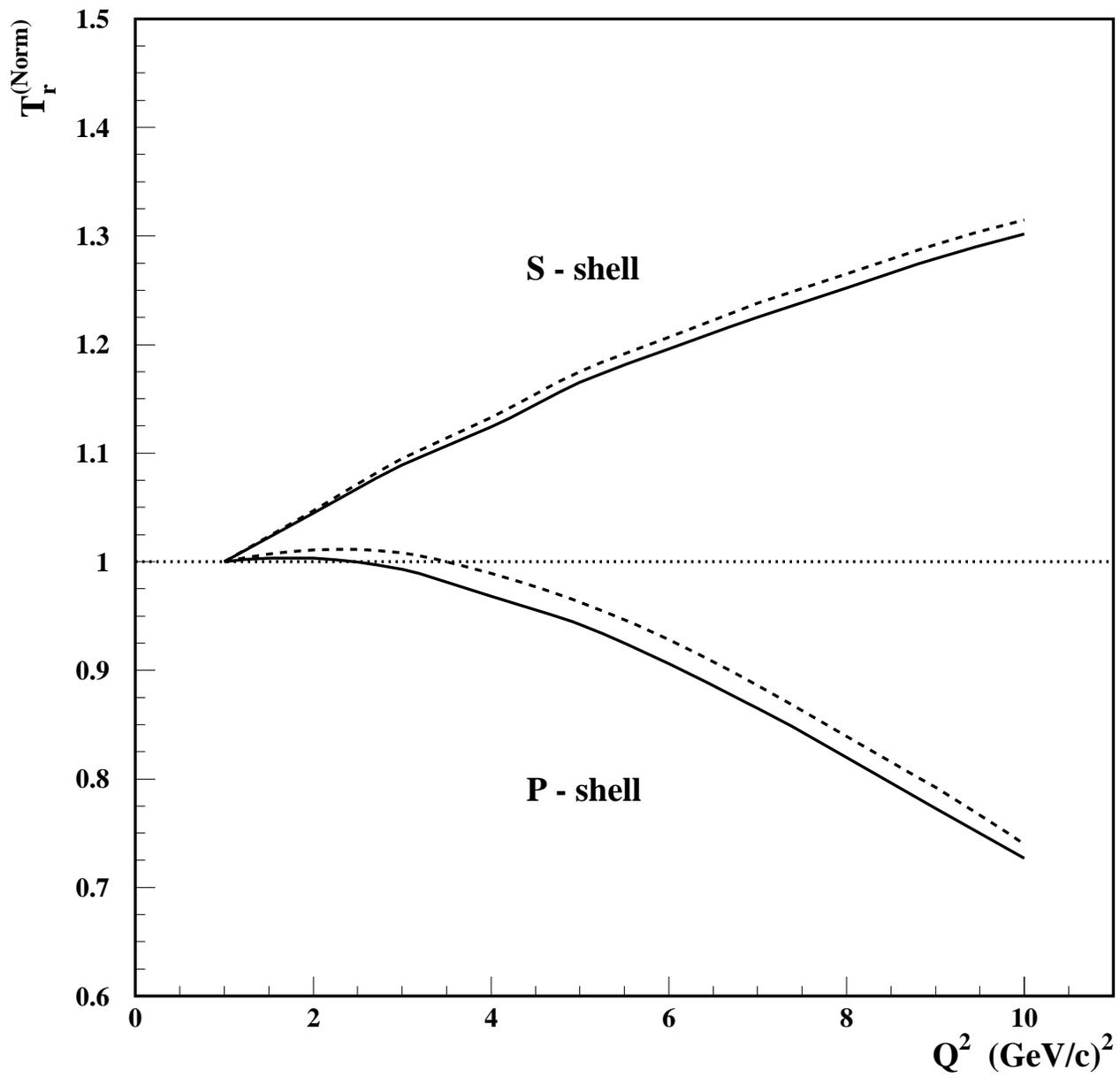
\item [ Figure 4.]  The $Q^2$ dependence of color transparency effect
normalised to the corresponding transparency within Glauber approximation.
Curves labeled  $s-shell$, $p-shell$ corresponded to the fixed shell
scattering, with coherent FSI, from $s$-shell, $p$-shell. For all
cases $\vec p_i~=~0$.  Dashed line - is the calculation without
correlation effects, solid line - with correlation effects.

\end{itemize}

\end{document}